\documentclass[12pt,preprint]{aastex}
\usepackage{epsf}

\begin{document} 

\title{Prospects for Detection of Catastrophic Collisions in Debris Disks}

\author{Scott J. Kenyon}
\affil{Smithsonian Astrophysical Observatory, 60 Garden Street, 
Cambridge, MA 02138, USA; e-mail: skenyon@cfa.harvard.edu}
\author{and}
\author{Benjamin C. Bromley}
\affil{Department of Physics, University of Utah, 201 JFB, Salt Lake City, 
UT 84112, USA; e-mail: bromley@physics.utah.edu}

\begin{abstract}

We investigate the prospects for detecting dust from two body
collisions during the late stages of planet formation at 1--150 AU.
We develop an analytic model to describe the formation of a 
dusty cloud of debris and use numerical coagulation and $n$-body
calculations to predict observable signals from these events.  
In a minimum mass solar nebula, collisions of 100--1000 km objects
at distances of 3--5 AU or less from the parent star are observable 
at mid-infrared wavelengths
as bright clumps or rings of dust. At 24 $\mu$m, the clumps are 
$\sim$ 0.1--1 mag brighter than emission from dust in the background
debris disk. In edge-on systems, dusty clumps produce eclipses with 
depths of $\lesssim$ 1.0 mag that last for $\sim$ 100 orbital periods.
Large-scale surveys for transits from exosolar planets, such as
{\it Kepler}, can plausibly detect these eclipses and provide important
constraints on the terrestrial environment for ages of $\lesssim$
100--300 Myr.

\end{abstract}


\section{INTRODUCTION}

Throughout history, collisions have shaped the Earth.  Repeated 
mergers of 10 km and larger objects produced the proto-Earth
in 10--30 Myr \citep{ws93,agn99,cha01,yin02}. 
A few Myr later, a giant collision with a Mars-sized planet removed 
the first terrestrial atmosphere and led to the formation of the 
Moon \citep{har75b,cam76,ben86,can04a,can04b}.
During the Late Heavy Bombardment $\sim$ 4 Gyr ago, a sustained 
period of large collisions shaped the surfaces of all terrestrial 
objects \citep[e.g.,][]{ter74, har80, neu83, ryd02, koe03}. 
Giant collisions also produced structure in the Zodiacal dust band 
and perhaps mass extinctions on the Earth 
\citep[e.g.][]{alv80,nes02b}. 

Understanding the formation and early evolution of our solar system 
requires unraveling this collisional history from current images, 
orbits, and samples of asteroids, meteorites, planets, and satellites.  
Although these data yield a remarkably detailed picture of the 
formation of the solar system, the record is incomplete, particularly
at the earliest epochs 
\citep[e.g.,][]{swi93,mel93,wad00}. 
Some of the early collisional history is also erased by subsequent 
events, including volcanic activity and collisions 
\citep[e.g.,][]{wil87,str94}.
The ambiguity of the early record leads to uncertainties
in our understanding of
the formation of chondrites \citep[e.g.,][]{ito03,woo04,con05},
the origin of the Late Heavy Bombardment 
\citep[e.g.][]{coh00,lev01,koe03}, 
and other aspects of the early history of the solar system.

Observations of debris disks surrounding other stars provide 
another way to study collisions during the evolution of a 
solar system. At least 40\% of all A-type stars and 10\% of
all solar-type stars have far-infrared excesses from 
circumstellar dust \citep[e.g.,][]{bac93}.  Debris disks with 
masses of $\sim$ 0.01 $M_{\oplus}$ in small, 1--100 $\mu$m 
grains explain the scattered-light images and the spectral 
energy distributions of the best-studied systems 
\citep[e.g.,][and references therein]{bac93,art97,gre05a}. 
However, collisions and radiation forces remove small grains on
timescales, $\lesssim$ 1 Myr, much shorter than the 10 Myr to 1 Gyr age 
of the parent star \citep{aum84,bac93,zuc01,the03}. If the mass in 
1--10 km bodies is $\sim$ 10--100 $M_{\oplus}$, debris produced from 
high velocity collisions among these larger bodies can replenish the 
small grain population for 100 Myr or longer \citep{hab01,kb02,dom03}. 
Detailed numerical calculations suggest that the formation of debris 
disks coincides with the growth of 1000 km or larger planets throughout 
the disk on similar 10--100 Myr timescales \citep[e.g.][]{kb04a,kb04b}.  
Thus, debris disks also provide a signpost for the formation of solar
systems.

Recent theoretical work helps to link our understanding of the 
collisions in debris disks with the collisions that shaped our 
solar system.  For debris disks around A-type stars, \citet{dom03} 
and \citet{kb02,kb04b} showed that the ratio of the disk luminosity 
$L_d$ to the stellar luminosity $L_{\star}$ should follow a simple 
power-law relation, $L_d / L_{\star} \approx L_0 / (1 + t/t_0)$, 
where $L_0 \sim$ $10^{-3}$ to $10^{-2}$ and $t_0 \sim$ 100--300 Myr.  
Observations of infrared excesses in nearby stars follow
a similar relation \citep[e.g.][]{kal98,rie05,gre05b}.  These 
results suggest an exponential decline in the collision rate,
similar perhaps to the monotonic decline in the background
accretion rate following the formation of the Moon\footnote{We 
distinguish a general monotonic decline in the impact rate 
4--4.5 Gyr ago from the Late Heavy Bombardment, an apparent spike 
in the impact rate $\sim$ 600 Myr after the formation of the Earth 
and the Moon \citep[see, for example,][and references therein] 
{har75a,har80,ryd88,har00,ryd02,koe03}.}
\citep[e.g.,][]{ryd02,koe03}. 
Among others, \citet{wya02} noted that large collisions are frequent 
enough to produce dusty clumps in debris disks, with long-term 
structure similar to the local Zodiac 
\citep{nes02b}.  In the terrestrial zone, \citet{kb04a} demonstrated 
that collisions between 50--100 km and larger objects can raise 
the debris disk luminosity above the simple power-law relation,
suggesting that moderate and giant impacts are observable for
short periods of time \citep[see also][]{ste94,gro01,nes02a,zha03}.  

Here, we continue to develop the idea that giant collisions
produce observable structures in debris disks. We derive an
analytic model to identify collisions that produce observable clumps 
of dust and we use numerical calculations to explore the conditions 
where large collisions are visible.  In a minimum mass solar nebula, 
collisions of 100--1000 km objects at distances of 3--5 AU or less 
from the parent star are observable as bright clumps or rings of dust.  
In disks observed close
to edge-on, the optical depth is often large enough to produce 
eclipses of the parent star. These eclipses last for several days 
and recur over 10--100 orbital periods. We use detailed $N$-body 
simulations to illustrate the evolution of these eclipses and 
to make estimates for the frequency of these events. Current 
and planned surveys for planetary transits can detect these 
eclipses and test our predictions.

We develop the analytic model in \S2, describe numerical
simulations of an ensemble of colliding planetesimals in \S3, 
and derive the time evolution of single catastrophic collisions
in \S4. We conclude with a brief discussion and summary in \S5.

\section{ANALYTIC MODEL}

\subsection{Derivation}

We begin with an analytic model for the formation of a clump of 
dusty debris from two colliding planetesimals. Each planetesimal 
has mass density $\rho$.  The larger planetesimal is the `target' 
and has mass $m_t$, radius $r_t$, and orbital eccentricity $e_t$ 
and inclination $i_t$. The smaller `projectile' has mass $m_p$, 
radius $r_p$, eccentricity $e_p$, and inclination $i_p$. The
planetesimals lie in a disk with surface density $\Sigma(a)$,
where $a$ is the distance from a star with mass $M_{\star}$ and
luminosity $L_{\star}$.

We assume that the collision produces a spherical cloud of debris 
with radius $r_c$ and mass $m_d$.  The debris consists of particles
with sizes ranging from 1 $\mu$m to 1 m or larger. We follow previous 
investigators and derive the debris mass from the ratio of the 
impact energy $Q_I$ to the disruption energy $Q_d$,
\begin{equation}
m_d = 0.5 (m_p  + m_t) \left ( \frac{Q_I}{Q_d} \right )^{\beta_e} ~ ,
\end{equation}
where $\beta_e$ is a constant of order unity
\citep[see][and references therein]{dav85, ws93, kl98, ben99}.
If $V$ is the relative velocity of the two planetesimals and
$V_e$ is the escape velocity of a single body with mass, 
$m = m_p + m_t$, the center-of-mass collision energy is \citep{ws93}
\begin{equation}
Q_I = \frac{m_p m_t V_I^2}{4 m^2} ~ ,
\end{equation}
where the impact velocity $V_I^2 = V^2 + V_e^2$.
The energy needed to remove half of the combined mass of two 
colliding planetesimals is 
\begin{equation}
Q_d = Q_b \left ( \frac{r}{\rm 1 ~ cm} \right)^{\beta_b} + \rho Q_g \left ( \frac{r}{\rm 1 ~ cm} \right)^{\beta_g} ~ ,
\end{equation}
where $Q_b r^{\beta_b}$ is the bulk (tensile)
component of the binding energy and
$\rho Q_g r^{\beta_g}$ is the gravity component
of the binding energy
\citep[see, for example,][]{dav85,ws93,hls94,ben99,hou99}.

To derive the amount of stellar radiation intercepted by the debris 
cloud, we define the radial optical depth through the cloud as
\begin{equation}
\tau \approx \frac{3 \int n(r) r^2 dr}{ r_c^2 } ~ ,
\end{equation}
where $n(r)$ is the number of particles with radius $r$ within
the debris cloud.  
The radiation intercepted by the debris is roughly 
$L_d/L_{\star} \approx \tau r_c^2 / 4 a^2$. As orbital shear
broadens the debris cloud into a ring, $\tau$ drops but 
$L_d/L_{\star}$ remains roughly constant. If $n(r) = n_0 r^{-\alpha}$ 
and $ m_d = (4 \pi \rho /3) \int n(r) r^3 dr$, $L_d/L_{\star}$ 
is a function of $a$, $m_d$, and $\alpha$ and does not depend 
on the initial radius of the debris cloud $r_c$.

The collision rate places strict limits on $\alpha$. For the late 
stages of planet formation, the timescale for a collision between 
an object and smaller bodies in the disk is \citep{lis87,ws93}
\begin{equation}
\frac{t_c}{P} \approx \frac{\rho r}{\pi n \Sigma} ~ ,
\end{equation}
where $P$ is the orbital period.  If we integrate the expressions 
for $\tau$ and $m_d$ over $r$, adopt $\rho$ = 1--3 g cm$^{-3}$, 
and set the surface density of small particles in the background
debris disk to $\Sigma = \Sigma_b a^{-3/2}$ with $\Sigma_b \approx$ 
0.1 g cm$^{-2}$ at 1 AU, we can rearrange equation (5) as
\begin{equation}
\frac{t_c}{P} \sim 10^{-25} ~ r^3 \left ( \frac{{\rm 1 ~ AU}}{a} \right )^{1/2} \left ( \frac{L_{\star}}{L_d} \right ) ~ .
\end{equation}
In each orbit, particles with $t_c/P < 0.1$ have many collisions 
with material in the background debris disk. Because the background
has a wide range of sizes, $r \sim$ 1--10 $\mu$m to $r \sim$ 1 km, 
objects in the debris cloud rapidly reach an equilibrium size 
distribution with $\alpha \approx$ 3.5 \citep{doh69,dur02,wil94}. 
For an observable debris cloud with $L_d/L_{\star} \gtrsim 10^{-5}$, 
particles with $r \lesssim$ 10--20 km reach this limit at $a$ = 1 AU.  
Although larger particles may have a steeper size distribution
\citep{cam01,mic01,mic02}, these contribute little to the optical depth.
Thus, we assume $\alpha \approx 3.5 $ for all particles in the debris.

With $\alpha$ known, the optical depth and luminosity are

\begin{equation}
\tau \approx \frac{m_d}{ \pi \rho \langle r \rangle r_c^2 } ~ ,
\end{equation}
where $\langle r \rangle$ is a typical radius of a particle, and
\begin{equation}
\frac{L_d}{L_{\star}} \approx \frac{m_d}{4 \pi \rho \langle r \rangle a^2}.
\end{equation}
The reprocessed luminosity is independent of the geometry of the 
debris cloud and depends only on the location of the collision and 
the mass distribution of the particles in the debris.
For $L_d/L_0 \gtrsim 10^{-5}$, this expression constrains
the total debris mass:
\begin{equation}
m_d \gtrsim 10^{21} \left ( \frac{\langle r \rangle}{\rm 1 ~ mm} \right
)
\left ( \frac{a}{\rm 1 ~ AU} \right )^2 ~ \rm g.
\end{equation}
Thus, an observable debris cloud requires the complete disruption of 
(i) two 100 km objects at 1 AU, (ii) two 300 km objects at 10 AU, 
or (iii) two 1000 km objects at 100 AU. From equation (5) these 
debris clouds form every $\sim 10^2$ yr at $\sim$ 1 AU and every
$\sim 1-3 \times 10^6$ yr at 30 AU. 

The timescale for orbital shear to broaden the debris cloud into 
a ring and mix with the background disk is $t_m \approx (a / r_c ) P$.  
After the collision, the debris has a velocity dispersion\footnote{The 
range in the velocity dispersion is a factor of $\sim$ 2--3. After the
collision, dynamical interactions with other objects and radiative
processes such as the Yarkovsky effect can modify the velocity 
dispersion \citep[e.g.,][]{mic02,mic04}.}
$V_d / V_K \sim 10^{-3}$ \citep[e.g.][]{mic01,mic02}, 
which implies an approximate size 
$r_c / a \approx$ $V_d/V_K \sim$ $10^{-3}$.
Gravity sets a lower limit to the extent of the debris cloud.
Debris cannot remain bound to the largest objects beyond the 
mutual Hill radius, $R_H/a = (2 m_p/3 M_{\star})^{1/3}$
$\sim 1-4 \times 10^{-4}$ for 100--300 km projectiles.  The disk 
scale height, $H \approx$ 0.01-0.1$a$ provides a rough upper limit 
to the width of the debris cloud.  These estimates suggest that 
the cloud remains distinct for at least $10^2$ orbital periods.

Our approach ignores the increase in thermal energy during a 
collision.  From detailed numerical calculations 
\citep[e.g.,][]{can04b}, large impacts produce considerable increases 
in the temperature of the colliding bodies.  The results suggest 
that the temperature of two large bodies rises to $\sim$ 5000 K
to 10,000 K during the impact \citep[see also][]{ste94}. At these
temperatures, the timescale to radiate the impact energy is short,
$\sim$ days, compared to the lifetime of the debris cloud, $\sim$ 
years. If the increase in thermal energy is large compared to the 
impact energy, our analysis overestimates the amount of debris and 
therefore yields upper limits to the debris cloud luminosity.

\subsection{Application}

To apply the analytic model to debris disks, we use parameters 
appropriate for the late stages of planet formation.  We adopt a 
minimum mass solar nebula model for the surface density, 
$\Sigma = \Sigma_0 (a/a_0)^{-3/2}$, where $a_0$ = 1 AU
and $\Sigma_0$ = 15 g cm$^{-2}$ \citep{wei77, hay81}. For this
section, we ignore the change in surface density at the `snow line,'
where the surface density increases due to ice condensation.
This approximation decreases our collision rates by a factor of
two, but has no impact on the luminosity or optical depth of a
debris cloud.

The size distribution of projectiles and targets is \citep{kb04c}
\begin{equation}
n(r) = \left\{ \begin{array}{l l l}
        n_S r^{-\alpha_S} & \hspace{5mm} & r \le r_1 \\
\\
	n_I               & \hspace{5mm} & r_1 \le r < r_0 \\
\\
        n_L r^{-\alpha_L} & \hspace{5mm} & r \ge r_0 \\
         \end{array}
         \right .
\end{equation}
Numerical simulations of planet formation suggest $r_0 \approx$ 
1--100 km, with $\alpha_L \approx$ 4 for large objects and 
$\alpha_S \approx$ 3.5 for small objects 
\cite[e.g.][]{st97,kb04a}. We use $r_0 \approx$ 10-100 km and
$r_1 \approx$ 0.1-1 km \citep{kb04c}.  For the orbital parameters, 
we use $M_{\star}$ = 3 $M_{\odot}$, $e$ = 0.2, and i/e = 0.5.

We assume standard values for the bulk parameters of large objects. 
In our model, rocky objects have $\rho$ = 3.0 g cm$^{-3}$,
$Q_b$ = $6 \times 10^7$ erg g$^{-1}$, $\alpha_b$ = $-$0.37,
$Q_g$ = 0.4 erg cm$^{-3}$, and $\alpha_g$ = 1.36; icy objects have
$\rho$ = 1.5 cm$^{-3}$, $Q_b$ = $10^3$ erg g$^{-1}$, $\alpha_b$ = 0.0,
$Q_g$ = 1.4 erg cm$^{-3}$, and $\alpha_g$ = 1.25.  For this paper,
we ignore a weak relation between the disruption energy and the
impact velocity \citep{hou90, hou99, ben99}.

To explore the sensitivity of our results to the bulk properties,
we consider two alternatives.
For rocky objects, our simple approximation to the empirical $Q_d$ 
derived from the size distribution of asteroids in the solar system 
\citep{dur98} is 
$Q_b$ = $5 \times 10^4$ erg g$^{-1}$, $\alpha_b$ = 0.0,
$Q_g$ = $1.7 \times 10^{-8}$ erg cm$^{-3}$, and $\alpha_g$ = 2.4. 
As an approximation for icy rubble piles, we also consider
$Q_b$ = 1 erg g$^{-1}$, $\alpha_b$ = 0.0,
$Q_g$ = $10^{-3}$ erg cm$^{-3}$, and $\alpha_g$ = 1.25
\citep[e.g.,][]{lei00,lei02}.

To calculate the properties of debris clouds, we use the orbital
parameters and bulk properties to derive $m_d$ from equations (1)--(3).
For $n(r) = n_0 r^{-\alpha}$ and $\alpha$ = 3.5, 
$n_0$ = $ 3 m_d / (4 \pi \rho \int r^{3-\alpha} dr$). 
With $r_c / a \lesssim$ $10^{-2}$, equation (4) yields a 
lower limit for the optical depth. The cloud luminosity is then 
$L_d/L_{\star} \approx \tau r_c^2 / 4 a^2$. Although $\tau$ depends
on $r_c$, the luminosity is independent of cloud geometry. 

Clouds viewed in the orbital plane of the disk can eclipse the central
star. The visual extinction through the cloud is $A_V = 1.086 \tau$. 
The maximum eclipse depth is 
\begin{equation}
\Delta m \approx A_V \left ( \frac{r_c}{a} \right )^2 \left ( \frac{a}{R_{\star}} \right )^2 ~ ,
\end{equation}
where $R_{\star}$ is the radius of the central star. At 1 AU, a debris
cloud with $r_c / a \approx 10^{-2}$ in orbit around a 1--3 $M_{\odot}$
star produces eclipses with $\Delta m \approx$ 0.1-1.0 $A_V$.

Figure 1 shows derived luminosities for debris clouds at 1 AU (rocky
objects, solid lines), 10 AU (rocky objects, dot-dashed lines), and 
100 AU (icy objects, dashed lines).  Larger projectiles produce more 
luminous debris clouds.  In some cases, the largest projectiles completely 
shatter the target, leading to a saturation in debris production. 
Complete disruption leads to a maximum in the relative 
luminosity for a given target, illustrated by the plateaus in the 
Figure. This maximum luminosity is roughly

\begin{equation}
\frac{L_{d,max}}{L_{\star}} \approx \left\{ \begin{array}{l l l}
	3 \times 10^{-2} & \hspace{5mm} & a \lesssim {\rm 1.6 ~ AU} \\
\\
	4 \times 10^{-5} \left ( \frac{a}{\rm 10 ~ AU} \right )^{-3.5} & \hspace{5mm} & a \gtrsim {\rm 1.6 ~ AU} \\
         \end{array}
         \right .
\end{equation}

The luminosity of a debris cloud is sensitive to the distance from 
the parent star.  Our results suggest a roughly 7 order of 
magnitude decline in $L_d/L_{\star}$ from 1 AU to 100 AU. This large 
change is a result of the changing angular scale and amount of debris 
with semimajor axis. For a cloud of fixed mass, the angular size of 
an opaque cloud as viewed from the central star varies as $a^{-2}$.  
For a fixed eccentricity, the 
impact kinetic energy scales as $a^{-1}$; thus, the debris mass produced 
from a collision of two bodies varies as $a^{-1}$.  Because collisions
at larger stellar distance produce less debris, the mass of the largest 
remnant increases roughly with $a$. Taken together, these simple scaling 
laws suggest $L_d /L_{\star} \propto a^{-4}$.  Complete disruption becomes
more important close to the central star, reducing the scaling to 
$L_d /L_{\star} \propto a^{-3.5}$

The luminosity of the debris is also sensitive to the bulk properties of 
the colliding bodies (Figure 2). For collisions between the largest
objects at 1--10 AU, the $Q_d$ relation derived from asteroids in our 
solar system yields a factor of ten smaller luminosity as the `standard' 
relation for basalts from \cite{ben99}. The debris luminosity from
smaller collisions is unchanged from the standard relation. For the
\citet{dur98} $Q_d$ relation, small changes, $\pm$ 0.1, in $\beta_g$ 
yield factor of 3 variations in the debris luminosity.  At 100 AU, 
the $Q_d$ relation for rubble piles results in much more luminous debris 
clouds for collisions between 100--1000 km objects.  However, the 
luminosity is still small, $L_d/L_{\star} \lesssim 10^{-6}$, compared 
to typical luminosities, $\sim 10^{-5}$ to $10^{-3}$, of known debris disks.

Close to the central star, debris clouds can produce modest eclipses 
(Figure 3). At 1 AU, collisions between 300 km and larger objects 
have $A_V \sim$ 0.01--0.1 mag. Although smaller collisions produce
observable debris clouds in scattered or thermal emission, they
produce tiny eclipses with $\Delta m \lesssim 10^{-3}$ mag.
Beyond $\sim$ 3 AU, eclipses from all debris clouds are unobservable
with current techniques. 

Our estimates of the collision times indicate that individual 
collisions between large objects are observable 
\citep[see also][]{ste94,gro01,zha03}. At 1 AU, the collision rates range 
from 1--10 Myr$^{-1}$ for collisions between two 1000 km objects to
10$^{-2}$ yr$^{-1}$ for 100 km + 100 km collisions. Smaller collisions 
occur more frequently but do not produce significant eclipses or
debris luminosity. At 10--100 AU, observable collisions between 300 km
and larger objects occur every 10 Myr to 1 Gyr. Because the debris
cloud merges with the disk on $10^4$ to $10^5$ yr timescales, the
chances of observing these debris clouds are small.

\subsection{Uncertainties}

The main uncertainties in our estimates are collision physics
and the behavior of the debris cloud following the collision
\citep[e.g.,][]{dur04}.
Because collisions between small objects are frequent, the size 
distribution of small bodies probably is a small uncertainty.
For optically thin clouds, the cloud geometry sets the optical
depth of the cloud but not the luminosity.

For rocky objects, the collision parameters have been well-studied
with numerical simulations \citep[e.g.,][]{ben99,mic01,can04a} and
observations of the debris of large collisions among asteroids
\citep[e.g.,][]{dur98}. The two approaches yield similar optical
depths and cloud luminosities for $r \lesssim$ 100 km, but differ
by a factor of $\sim$ 10 for objects with $r \sim$ 1000 km. Because 
the number of large asteroids with $r \gtrsim$ 500 km is small, the 
uncertainty in the observational estimates are large \citep{dur98}. 
Observations of sufficient numbers of collisions in debris disks may 
reduce the uncertainty (see below).

For icy objects, the uncertainties in the bulk properties are much 
larger. Numerical simulations yield $Q_b \sim 10^6$ erg g$^{-1}$ 
for crystalline ice or snow \citep{ben99}; simulations of rubble 
piles suggest $Q_b \lesssim$ 10 erg g$^{-1}$ \citep{lei02}.  The 
rubble pile results agree with analyses of simulations of the break 
up of comets, such as Shoemaker-Levy 9 \citep[e.g.,][]{asp96}
and C/2000 C6 \citep{uzz01}. 
The apparent break in the size distribution of Kuiper belt objects 
in the outer solar system \citep{ber04} also favors 
$Q_b \lesssim 10^3$ erg g$^{-1}$ \citep{pan05,kb04c}.

The size distribution of Kuiper belt objects also favors a small
exponent for the gravity component of the disruption energy,
$\beta_g \lesssim$ 1.2--1.3 instead of the $\beta_g \approx$ 2 
expected from simple theory \citep{kb04c}. However, current observations
cannot distinguish between the two relations used in Figures 1--3.  
Improved constraints from the size distribution of Kuiper belt 
objects may yield tests of these relations \citep{ber04}.

The evolution of the debris cloud following the collision is an
important uncertainty in our estimates.  For a cloud with a radial
extent $r_c$, the number of secondary collisions with material in
the background debris disk ranges from
less than one per orbit for the largest remnant of the collision
to a hundred or more collisions per orbit for debris particles 
with radii of 1 km or smaller. Because the debris cloud expands
azimuthally at a rate roughly comparable to the collision rate,
the optical depth through the cloud probably cannot increase with
time. However, the mass and luminosity may increase. Depending on
the surface density in the disk, we estimate that secondary
collisions can increase the luminosity of the debris cloud by a 
factor of 2--10.  This uncertainty has little impact on our main 
results.  If the debris cloud luminosity is a factor of 10 larger
than our analytic estimates, individual collisions between objects 
with radii of 100 km or smaller are not detectable at 1 AU. At
10--100 AU, individual collisions between objects with radii of
1000 km and smaller remain invisible.

\section{Coagulation Calculations}

\subsection{Methods}

To explore the visibility of individual collisions in more detail, 
we consider numerical calculations of the evolution of particle disks 
around 1--3 $M_{\odot}$ stars.  Our calculations for the terrestrial 
zone around solar-type stars show that individual collisions between 
100 km and larger objects are visible as short-lived increases in 
the disk luminosity \citep[][see also Grogan et al. 2001]{kb04b}.
Beyond 30 AU, individual collisions are not visible as discrete events 
in our calculations \citep{kb04a}. Here, we consider a broader suite 
of calculations designed to test the visibility of individual collisions 
at 1--100 AU. 

To evolve a particle disk in time, we follow the \citet{saf69} 
statistical approach to calculate the collisional evolution of an 
ensemble of planetesimals in orbit around a star of mass $M_{\star}$ 
and luminosity $L_{\star}$ \citep[][and references therein]{kb04c}.
The model grid contains $N$ concentric annuli with widths $\delta a_i$
centered at heliocentric distances $a_i$.  Calculations begin with
a differential mass distribution $n(m_{ik}$) of bodies with
orbital eccentricity $e_{ik}(t)$ and inclination $i_{ik}(t)$.

To evolve the mass and velocity distributions in time, we solve 
the coagulation and Fokker-Planck equations for bodies undergoing 
inelastic collisions, drag forces, and long-range gravitational 
interactions \citep{kb02}.  We adopt collision rates from kinetic
theory -- the particle-in-a-box method -- and use an energy-scaling
algorithm to assign collision outcomes \citep{kb04c}.  We derive 
changes in orbital parameters from gas drag, dynamical friction, 
and viscous stirring \citep{ada76, oht02}.  Dynamical friction 
transfers kinetic energy from large bodies to small bodies and 
drives a system to energy equipartition.  Viscous stirring transfers 
angular momentum between bodies and increases the velocities of 
all planetesimals.  

To compute dust masses and luminosities, we use the dust production
rates and particle scale heights derived from the coagulation code 
as input to a dust evolution code \citep{kb04a}. This calculation 
includes collisions between dust grains and Poynting-Robertson drag.
We use a simple radiative transfer method to derive the optical
depth through the disk; dust luminosities follow from the scale
height derived from the coagulation code.

Our numerical calculations begin with 1--1000 m planetesimals in
32--64 concentric annuli with initial surface density 
$\Sigma = \Sigma_0$ (a/1 AU)$^{-3/2}$ 
\citep[see, for example,][]{kl99,kb04a,kb04b}. 
Table 1 lists $a_{in}$ and $a_{out}$, the inner and outer radii 
of the computational grid, 
$\rho$, the mass density of a planetesimal,
$\Sigma_0$, and $t_g$, the timescale for exponential gas removal
\citep{kl99}.
We divide the initial continuous mass distribution of planetesimals 
into a differential mass distribution with discrete mass batches. 
The spacing between successive mass batches is $\delta = m_{i+1}/m_i$.
We add mass batches as planetesimals grow in mass.  Each planetesimal 
batch begins in a nearly circular orbit with eccentricity $e_0$ and 
inclination $i_0 = e_0/2$.  For these calculations, $\delta$ = 1.4--2
and $e_0 = 10^{-5}$--$10^{-4}$.

In the next two sections, we consider the evolution of icy and
rocky planetesimals in orbit around solar-type and A-type stars.
Because our calculations do not include gas accretion, we avoid
the region of the gas giant planets and restrict our attention 
to the terrestrial zone and the region of the Kuiper belt around 
solar-type stars. For A-type stars, we consider two representative 
regions at 3--20 AU and at 30--150 AU. Although gas giants may form 
around A-type stars, our calculations provide reasonable estimates 
for the formation of the rocky or icy cores of gas giants. As a
planetary core accretes gas from the disk, it will continue to
stir up the leftover planetesimals and remove the debris more
rapidly than estimated in our calculations.
Thus our estimates for the lifetimes and dust luminosity of
debris disks are probably reasonable upper limits for these systems.

\subsection{Solar-type stars: results at 0.4--2 AU and 30--150 AU}

At 1 AU, the growth of planetesimals into planets follows a standard 
pattern \citep{ws93,wei97}. All bodies begin with small orbital 
eccentricities. Gas drag maintains small $e$ for the smallest bodies;
dynamical friction maintains small $e$ for the largest bodies. 
Thus, collisions produce larger bodies instead of smaller fragments.
After $\sim 10^4$ yr, 1000 km and larger bodies form and stir up the
orbits of planetesimals with radii of 0.1--10 km. Collisions between 
the leftover planetesimals then produce smaller fragments.  The 
largest bodies sweep up some of the fragments, but most of the 
fragments collide with other fragments to produce still smaller 
fragments.  Because the largest bodies are nearly immune to 
fragmentation, this collisional cascade concentrates most of the 
initial mass in the largest and the smallest bodies.

During the earliest stages of our calculations, the collisional cascade 
produces an observable amount of dust (Figure 4). After $\sim 10^3$ yr,
the dust produces an order of magnitude more radiation at mid-infrared 
(mid-IR) wavelengths than the central solar-type star. Continued dust 
production maintains a roughly constant mid-IR excess for $\sim 10^6$ yr.
Because the large bodies are slowly depleted, the dust production rate 
then slows and the mid-IR excess declines. In these calculations, the
mid-IR excess becomes comparable to the stellar radiation after $\sim$
a few $\times ~ 10^7$ yr. 

In addition to the slow evolution of the `background' mid-IR excess,
individual collisions produce large spikes in the dust luminosity 
(Figure 4). During the early stages of the evolution, collisions
between large objects tend to produce even larger objects and are 
also too frequent for any single, disruptive collision to modify 
the mid-IR excess significantly.  As the average dust production 
rate declines, individual collisions between large objects are less 
frequent but produce more dust. Thus, after $\sim 10^6$ yr, spikes 
from large collisions dominate the evolution of the mid-IR luminosity.

To learn more about the nature of these spikes, we consider the dust 
production rate as a function of time (Figure 5). At this point
in the evolution ($\sim 5 \times 10^6$ yr), numerous collisions between
small objects set the background dust production rate. Because we
use a random number generator to assign collision outcomes
\citep{ws93,kl98}, an
occasional lack of collisions leads to a smaller than average dust 
production rate.  In addition, occasional collisions between
larger objects produce distinct spikes in the dust production rate.
The frequency distribution of the spikes in the dust production
rate tracks the size distribution of large bodies; small spikes
are much more frequent than large spikes.  The smaller spikes, 
where the dust production rate is a few times the background rate, 
produce little change in the mid-IR excess. Larger spikes -- produced
by collisions between 100 km or larger objects -- yield fluctuations
in the mid-IR excess ranging from a few percent to more than 100\%. 
The largest fluctuations in Figure 5 produce the largest spikes in
the evolution of the mid-IR excess shown in Figure 4.

At 30 AU, the evolution follows a different path.  Because the orbital 
periods are longer and the surface density smaller, planets take much 
longer to form at 30 AU than at 1 AU. Large bodies begin to form at 
$\sim$ 10 Myr and soon contain a few percent of the initial mass. 
These objects stir up leftover planetesimals along their orbits and 
begin the collisional cascade. In these models, large bodies grow more 
slowly, and the leftovers are more easily shattered. Collisions then
efficiently convert most of the initial mass into smaller objects which 
are removed by radiation pressure and Poynting-Robertson drag. Thus,
the final ensemble of large bodies contains less than 5\%, and sometimes 
less than 1\%, of the initial mass in solid objects.

The evolution of the IR excess is also different at 30 AU than at 1 AU
(Figure 6).  Because the dust is farther away from the central star, 
the dust is
cooler, $\sim$ 30--40 K instead of $\sim$ 300--400 K. At the start of 
our calculations, this dust produces a significant IR excess, which
slowly declines as small bodies merge into larger bodies. Once the
collisional cascade begins at $\sim$ 10 Myr, the far-IR excess rises
significantly. At shorter wavelengths, the dust produces a small excess. 
As in calculations at 1 AU, the cascade maintains a roughly constant
excess until planet formation reaches the outer edge of the disk at
100--150 AU. Poynting-Robertson drag and radiation pressure then remove
small grains faster than collisions can replenish them. The IR excess
then rapidly declines. However, individual collisions never contribute
a significant amount to the IR excess. In our calculations, the 
evolution of the IR excess is smooth, with 1\%--2\% fluctuations
about the general trend.

In our calculations at 30 AU, there are several reasons for the lack
of significant spikes in the IR excess. Because the collisional 
cascade is efficient at removing 1--10 km bodies from the evolution,
few of these objects grow into 100 km and larger objects. Collisions
between individual objects also do not produce an observable amount
of dust (Figs. 1--2). To test the relative importance of these processes,
we calculated several models with more robust planetesimals, where
the collisional cascade leaves more material in larger bodies. These
calculations suggest that individual collisions never produce observable
amounts of dust. 

Our suite of calculations confirms the basic conclusions of our 
analytic model.  At 30 AU, binary collisions never produce an 
observable IR excess.  Although the dust in our calculations 
produces a large far-IR excess, the evolution of the excess is
very smooth.  At 1 AU, occasional collisions between large objects 
produce observable changes in the mid-IR excess. Small, few percent, 
changes in the mid-IR excess should recur on timescales of tens to 
hundreds of years. Large changes are much less frequent and recur 
on thousand year or longer timescales. 

\subsection{A-type stars: results at 3--20 AU and 30--150 AU}

Calculations for planet formation around A-type stars follow the 
same pattern as for solar-type stars. During the early stages,
runaway growth leads to the formation of protoplanets with 
radii of 1000 km and larger.  At 3 AU, it takes about $10^5$ yr 
for 1000--2000 km objects to form. Due to the large gradient in 
surface density, the growth of large bodies slowly propagates to 
larger distances \citep{lis87}. The timescale for the formation
of 1000 km objects is $\sim 10^6$ yr at 9 AU and $\sim 10^7$ yr
at 18--20 AU. 

Once large planets form, their gravity stirs up the leftover
planetesimals along their orbits. Collisions then produce 
fragments instead of mergers, and the collisional cascade begins. 
In simulations at 3--20 AU, the dust luminosity starts to 
rise at $\sim 10^4$ yr and peaks at $\sim 10^6$ yr (Figure 7). 
As planet formation moves from $\sim$ 10 AU to $\sim$ 20 AU, 
the dust produced in the collisional cascade has smaller optical 
depth and subtends a smaller solid angle as seen from the central 
star. Thus, the dust luminosity declines. Once planet formation
reaches the outer edge of the 20 AU disk, Poynting-Robertson
drag removes dust more rapidly than collisions can replenish it.
The dust luminosity then declines more rapidly and falls to very
small levels in $\sim 10^9$ yr.

In calculations at 30--150 AU \citep{kb04a}, planet formation leads 
to dust production on much longer timescales. In these models, large 
planets begin to form at the inner edge of the disk at 10--20 Myr. 
After 1 Gyr, planet formation reaches 
the outer edge of the disk. Throughout this period, the collisional 
cascade maintains a large reservoir of small dust grains, which
produce a measurable dust luminosity (Figure 7). From 10 Myr to
1 Gyr, the dust luminosity smoothly declines from 
$L_D/L_0 \sim 10^{-3}$ to $L_D/L_0 \sim 10^{-4}$. After 1 Gyr,
Poynting-Robertson drag rapidly removes dust and the dust luminosity
declines more rapidly.

Figure 8 shows the evolution of the 25 $\mu$m excess for a planetesimal 
disk around an A-type star with mass $M_{\star}$ = 3 $M_{\odot}$, 
luminosity $L_{\star}$ = 50 $L_{\odot}$, and 
temperature $T_{\star}$ = 9500 K. 
In both cases, the excess first declines with time 
as small bodies merge into larger and larger bodies. Once the
collisional cascade begins, the 24 $\mu$m excess rises. Small
fluctuations from individual collisions mark the rise of the
excess at 3 AU. At 30 AU, the rise is smooth. At the peak of both 
calculations, small fluctuations in the excess are apparent. 
At 3 AU, individual collisions produce the small fluctuations.
At 30 AU, the occasional merger of large objects produces more
stirring and a temporary increase in the dust production rate. 

These results also confirm the basic predictions of the analytic
collision model. In planetesimal disks surrounding A-type stars,
binary collisions at 3--5 AU produce modest, $\sim$ 10\%, 
fluctuations in the mid-IR excess. Farther out in the disk,
binary collisions do not produce enough dust to change the 
amount of mid-IR excess. However, mergers of larger objects 
can produce modest fluctuations in the mid-IR excess by increasing
the local stirring rate, which leads to more dust and a larger 
mid-IR excess.

\section{N-BODY CALCULATIONS}

When binary collisions occur, our model predicts two observable 
changes in the luminosity. After a collision produces a debris cloud, 
we expect a rapid rise in infrared brightness. As orbital shear 
spreads the debris cloud into a ring, secondary collisions with 
material in the rest of the disk remove smaller particles. The 
mid-infrared flux then declines.  In our picture, this decline 
occurs on the timescale for the cloud to spread into a ring,
$\sim 10^2$ orbital periods.

For edge-on systems, we expect the debris cloud to eclipse the central 
star. The predicted eclipse depth ranges from $\sim$ 0.5 mag for large 
collisions to less than 0.01 mag for smaller collisions. As the debris 
cloud spreads into a ring, the eclipses should last longer and become
shallower. This behavior distinguishes a debris cloud from the transit
of an exosolar planet, where the eclipse depth and duration are
constant \cite[see, for example,][and references therein]{alo04}.

To illustrate the appearance of eclipses from a debris cloud, we use
$N$-body simulations to follow the evolution of debris following a
major two body collision.  For these examples, we consider collisions
between equal mass bodies with radius 300~km and 1000~km.

To begin the evolution, we place 10,000 ``dust'' particles in a small 
volume of radius $r_c/a = 10^{-3}$ around a guiding center on a 
circular orbit around a star of mass $M_{\star}$ = 1 $M_{\odot}$. We
give each dust particle a relative velocity $V/V_K \sim 10^{-3}$.  In 
one configuration, the positions and velocities have Gaussian random 
distributions (generating a ``random blob''). In another configuration,
the particle density declines as $r^{-2}$ out to radius $r_c$, 
with velocities proportional to the distance from the guiding center 
(yielding a homologous ``explosion''). These two configurations
represent two reasonable choices for the initial velocity distribution
\citep[e.g.,][]{mic01}.

Each $N$-body particle represents a population of dust with mass
distributed according to equation (10), with $\alpha = 3.5$. The
minimum size of the dust particles is 1~$\mu$m. We set the maximum
size in one of two ways. In the first case, we assume that all 
of the mass in the collision goes into ``dust particles'' with a 
single-index power law size distribution and a single large remnant 
containing 10\%--15\% of the mass. In the second model we suppose
that 20\% of the mass ends up in a single large object, with the
remaining 80\% in dust with a maximum radius of 10~km.  Although 
these choices do not exhaust the possible mass configurations
following a completely disruptive collision, they provide a good
illustration of the range of eclipse depths and evolution with time.

To derive the properties of the eclipses, we calculate the position 
of the particles on Keplerian orbits evaluated with an $8^{\rm th}$-order 
ODE solver.  We divide the surface of a star with radius 1 $R_{\odot}$ 
into $O(100)$ elements and project the particle positions onto the
stellar disk. For each path from the observer through the particles
to the stellar disk, we compute the optical depth $\tau$(t) and the
emergent intensity $I^{\prime}_i(t) = e^{-\tau(t)} I_i$, where the 
unobscured intensity from the star is $I_0 = \sum I_i$.  This procedure 
assumes that the mass distribution for all particles is identical. 
We then derive a light curve from -2.5 log $(\sum I^{\prime}_i(t))/I_0$.

Figure 9 illustrates light curves for the collision of two 1000 km
objects. The lower (upper) panel shows results for a random blob 
(explosion) when a large remnant contains 20\% of the initial mass.
In both simulations, the eclipse depth is $\sim$ 0.5--0.7 mag $\sim$ 
6 months after the collision.  Each eclipse lasts $\sim$ 5--10 days, 
$\sim$ 2\% to 4\% of the orbital period.  
After 10--20 orbits, the eclipses are $\sim$ 10\% of their original depth
and last $\sim$ 10 times longer than eclipses immediately after the
collision.  After $\sim$ 100 orbits, dust from both the blob and the
explosion lies in a roughly uniform ring around the star. Although
tiny eclipses remain visible, most of the dust produces a uniform 
extinction of $\sim$ 2\% to 3\% of the stellar flux.

In these simulations, dust from the random blob expands more rapidly
and merges into the background debris disk faster than dust from
the explosion. This behavior is a feature of the initial conditions.
In the homologous explosion, the dust particles are less concentrated
to the center, which produces shallower eclipses relative to the random
blob. However, particles close to the center of the explosion expand
more slowly than those in the blob.  Thus, these particles remain
closer to the point of the collision and continue to produce narrow, 
shallow eclipses when most of the rest of the blob has expanded into 
a ring.

Figure 10 repeats Figure 9 for the collision of two 300 km bodies.
Because these collisions produce less dust, the eclipses have roughly
the same duration but are shallower. As for the collision of 1000 km
bodies, the eclipses slowly fade away and last a longer and longer
fraction of the orbital period. After 30--100 orbits, orbital shear
transforms the debris cloud into a ring, which obscures $\sim$ 1\%
to 2\% of the stellar surface.

Figure 11 compares results for configurations where we place all of
the debris in ``dust'' particles with radii of 10 km or smaller.
The lower (upper) panel shows light curves for the collision of
two 1000 km (300 km) bodies.  Compared to configurations where 20\% 
of the initial mass in contained in a single large object (Figures 9--10), 
the optical depth in the debris cloud is larger and eclipses are much 
deeper. The timescale for the cloud to expand into a ring is 
identical to the timescale derived for the other cases. Thus, the 
magnitude of the eclipses declines over 30--100 orbital periods.

These calculations illustrate the simplest evolution of a debris
cloud formed in a large binary collision. Calculations with more
complex initial configurations produce a remarkable variety of eclipse 
light curves. Shepherd moons or planets can confine the debris cloud 
and thus can slow the decline of the eclipse depth with time.
Planets in orbits inclined to the orbital plane of the debris cloud
lead to a precessing cloud or ring with very different eclipse 
morphology. \citet{win04} and \citet{chi04} show how the eclipse 
morphology changes when an opaque cloud or ring orbits an eccentric 
binary star and apply their results to the young variable star 
KH 15D \citep{ham01}.
We derive similar results when a debris cloud spreads into a ring.  
Because the range of possible geometries is large, the discovery of 
additional eclipsing systems with rings or blobs of dust will guide 
future numerical calculations.

\section{DISCUSSION AND SUMMARY}

During the late stages of planet formation, binary collisions between 
leftover, large planetesimals in orbit around solar-type stars are
potentially observable.  At 1 AU, these collisions produce large spikes, 
$\sim$ 1 mag or less, in the mid-infrared luminosity (Figure 4). 
In systems viewed edge-on, the debris cloud eclipses the parent star 
(Figures 9--11). 
Maximum eclipse depths range from $\sim$ 0.5 mag for collisions between 
1000 km objects to $\sim$ 0.01 mag or less for collisions between 
100--300 km objects. Typical eclipses last 5--10 days. As the debris
cloud expands into a ring, the eclipses fade and grow longer. The 
lifetime of the debris cloud as a distinct entity is $\sim 10^2$ 
orbital periods.

At large distances from the parent star, $\gtrsim$ 3 AU, binary collisions
are more difficult to detect. Because collisions are less frequent 
and produce less dust, they are harder to detect against the emission
from the background debris disk. Our results indicate that the maximum
luminosity from individual collisions declines as $a^{-3.5}$ (equation
12). \citet{wya02} derive similar conclusions from different approaches.

To estimate the likelihood of detecting binary collisions, we consider 
{\it Kepler}\footnote{http:$//$www.kepler.arc.nasa.gov}, a satellite 
designed to detect the transits of exosolar planets across their
parent star. During a 4 yr mission, the {\it Kepler} team plans to 
acquire very high precision photometry for $10^5$ stars. Roughly 
0.5\% of these stars have favorable orbital inclinations to detect 
transits from terrestrial planets. For the planned observing frequency 
of once every few hours and an expected precision of 0.01\% or better, 
the expected detection rate is $\sim$ 50--100 terrestrial planets
per $10^5$ stars.

To modify this estimate for the detection of debris clouds, we
use the results from our coagulation and $n$-body calculations.
Because the scale height of the debris disk is larger than the 
radius of a terrestrial planet, $\sim$ 2\% to 3\% of potential 
solar systems have orientations that produce eclipses.  If the 
lifetime of the collision phase is a few hundred Myr, $\sim$ 10\% 
of {\it Kepler} targets are young enough to observe collisions.  
Thus, eclipses from debris clouds are possible in $\sim$ a few 
hundred stars of the {\it Kepler} sample. 

To estimate the probability of detecting binary collisions in
these stars, we adopt the collision rates from the coagulation
calculations and the lifetimes derived from the $n$-body results.
For the anticipated precision of {\it Kepler}, binary collisions between 
100 km objects yield a detectable eclipse.  With a collision rate 
of $10^{-2}$ yr$^{-1}$ for these objects, the probability of a
collision during the {\it Kepler} lifetime is roughly 100\%.
In a 4 yr mission, the eclipse depth can change by a factor of 
two or more. The duration of each eclipse lengthens to maintain 
a roughly constant total optical depth, integrated over the 
complete eclipse.

The prospect for observing collisions between the largest objects
is poorer but still significant.  The rate for collisions between 
two 1000 km objects is $\lesssim 10^{-5}$ yr$^{-1}$. This rate yields
a $\sim$ 1\% probability for detecting one large collision during the
{\it Kepler} mission.  However, with a typical lifetime of 30--100 yr, 
the probability of observing eclipses after a large collision is 
$\sim$ 10 times larger than observing a collision during the mission. 
Thus, it is possible that {\it Kepler} will detect eclipses from the 
aftermath of one large collision between two 1000 km objects. These 
eclipses should lengthen and decrease in depth throughout the 4 yr 
{\it Kepler} mission.

Ground-based surveys could also detect collisions in the terrestrial
zone. The {\it OGLE} and {\it TRES} projects have identified several 
transiting exosolar planets \citep{alo04,kon03,pon04}. The {\it PASS} 
project envisions a complete survey of planetary transits for all 
stars with V $\sim$ 5.5--10.5 \citep{dee04}. 
As part of a much deeper all-sky survey,
Pan-STARRS\footnote{http://pan-starrs.ifa.hawaii.edu/public/index.html}
might also detect eclipses from debris clouds.
To detect the unique signal from a large two body collision, these 
projects need to distinguish real events with diminishing depth and 
lengthening duration from spurious signals and from repetitive 
eclipses with constant depth and duration. 

Successful detections of the dust produced in large two body collisions 
can teach us about terrestrial planet formation. Characterizing the
frequency of events associated with stars of different ages and 
spectral types yields estimates for the space density of large
bodies as a function of stellar mass and age. Simultaneous optical,
infrared, and submm observations provide measures of the dust albedo
and size distribution.  Coupled with direct detection of exosolar
terrestrial planets, measurements of the properties of the terrestrial
environment allow a better understanding of the formation of the
Earth-Moon system and, ultimately, the formation of the current
environment on the Earth and other terrestrial planets.

\vskip 6ex

We acknowledge a generous allotment, $\sim$ 1000 cpu days, of
computer time on the Silicon Graphics Origin-2000 `Alhena' at the
Jet Propulsion Laboratory through funding from the NASA Offices of
Mission to Planet Earth, Aeronautics, and Space Science.  
T. Currie, M. Geller, J. Winn, and an anonymous referee 
provided helpful comments on the manuscript.  
The {\it NASA} {\it Astrophysics Theory Program} supported
part of this project through grant NAG5-13278.

\clearpage


\begin{figure}
\plotone{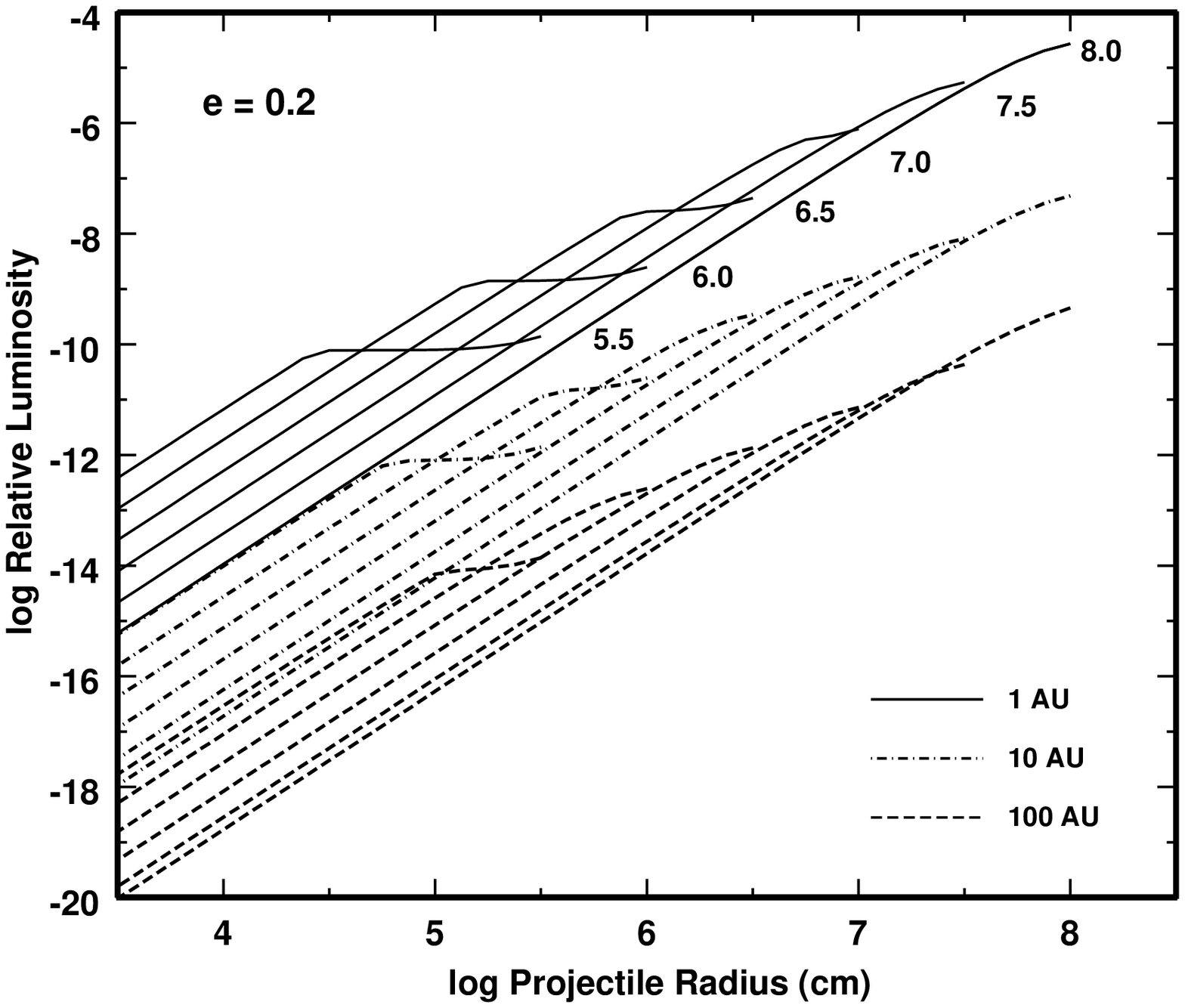} 
\caption
{Luminosity of debris clouds as a function of projectile radius at
1 AU (solid curves), 10 AU (dot-dashed curves), and 100 AU (dashed curves).
The log of the target radius (in cm) is listed to the right of each
solid curve. The target radii for the other curves follow the same
pattern. The curves assume bulk properties of 
(a) basalts from \citet{ben99} at 1--10 AU, and
(b) modified ices from \citet{ben99} as listed in the main text.
At 1 AU, collisions between 100 km and larger objects 
produce observable debris clouds. At 3 AU and beyond, individual
debris clouds are not visible.}
\end{figure}

\begin{figure}
\plotone{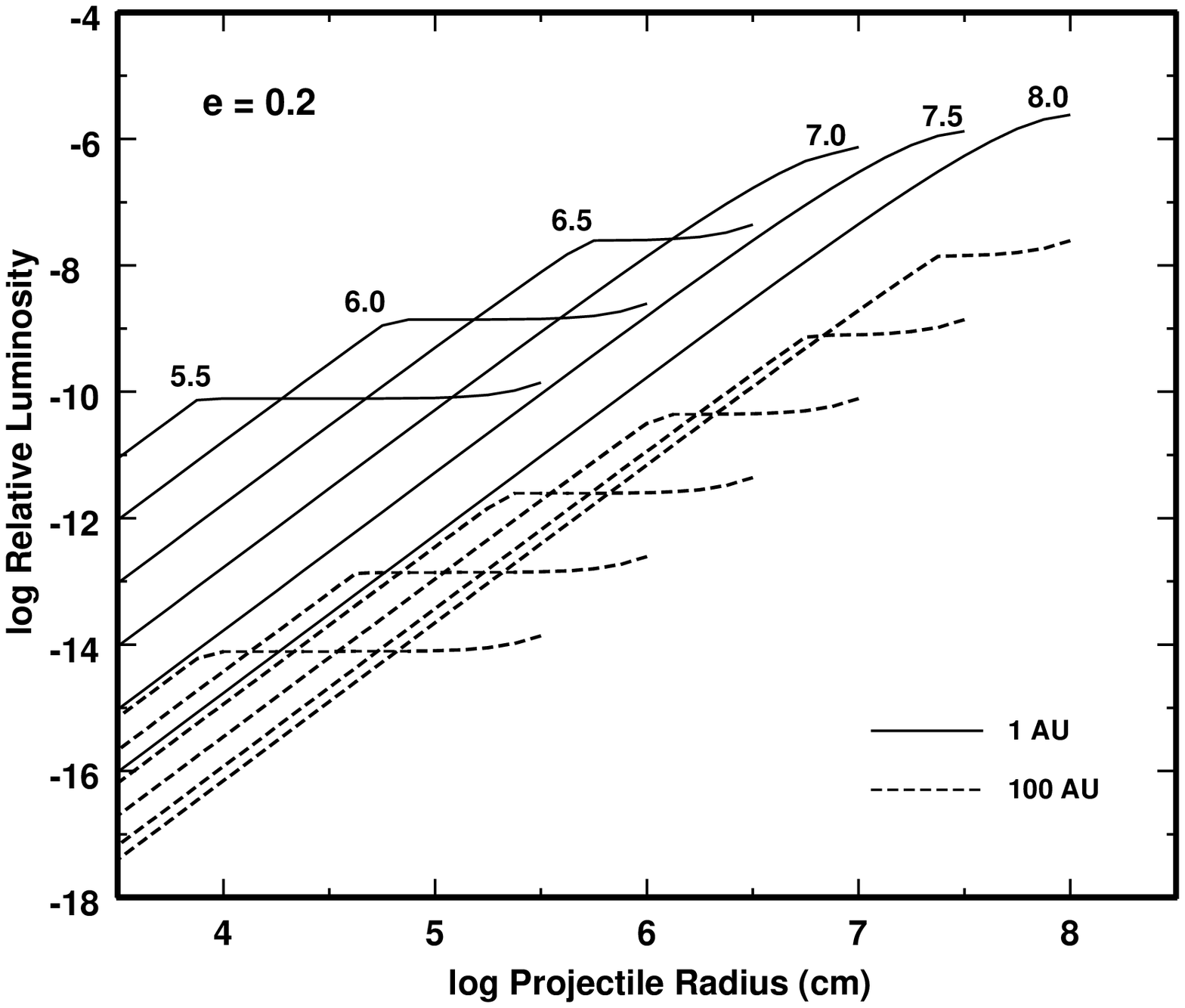} 
\caption
{As in Figure 1 for objects with 
(a) bulk properties for asteroids from \citet[][solid curves]{dur98},
and
(b) bulk properties of rubble piles from \citet[][dashed curves]{lei02}.}
\end{figure}

\begin{figure}
\plotone{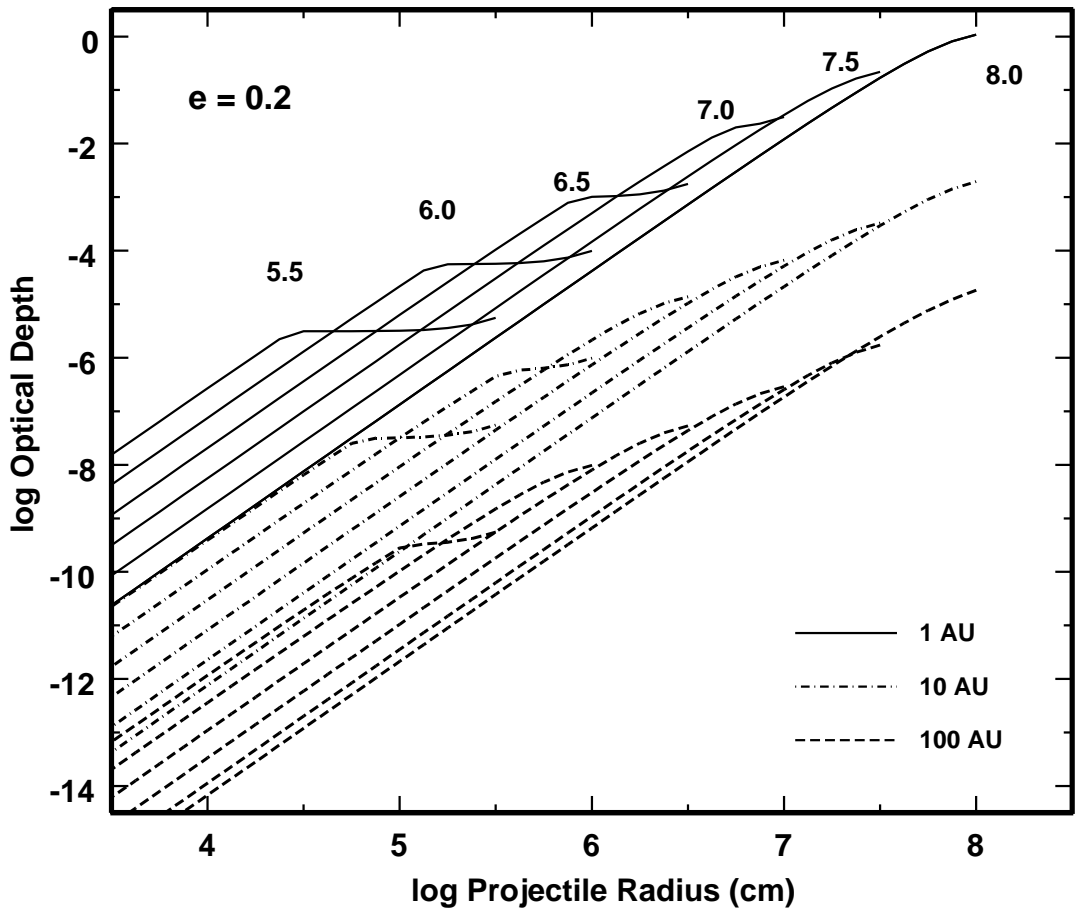} 
\caption
{Optical depths for the models in Figure 1.}
\end{figure}

\begin{figure}
\plotone{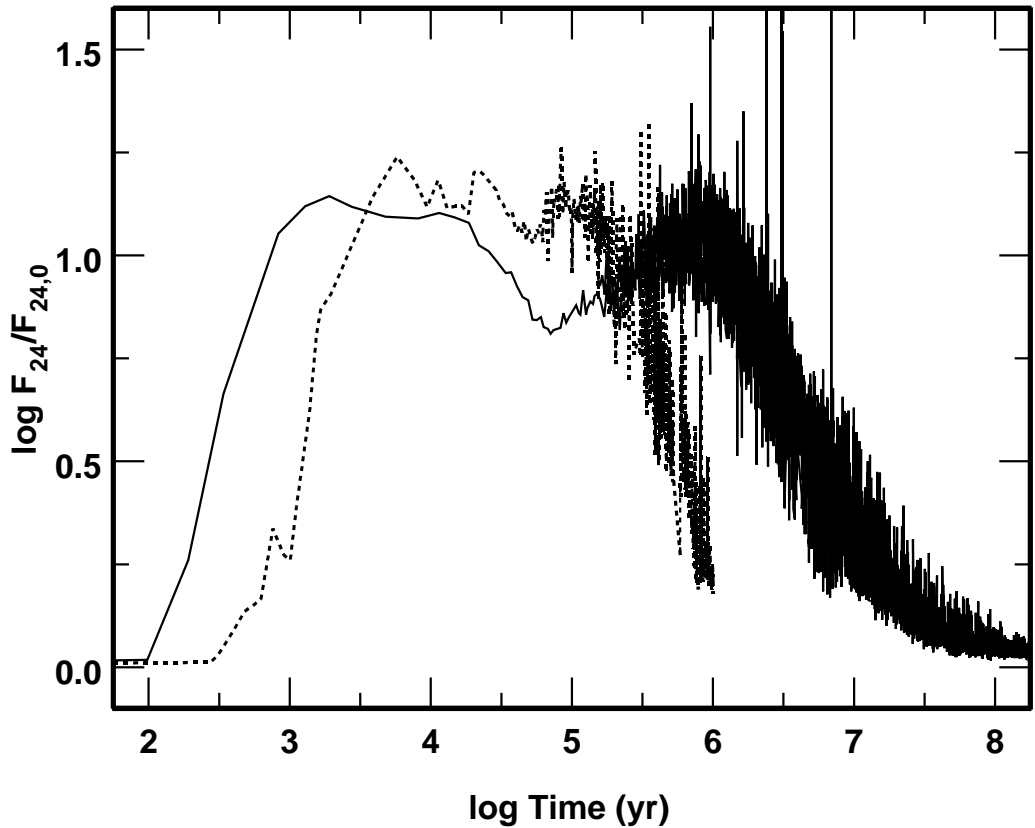} 
\caption
{Evolution of the 24 $\mu$m excess as a function of time
for planetesimal disks at 1.0 AU.  The radial extent of 
the disk is 0.68--1.32 AU (0.4--2.0 AU) for the dashed
(solid) curve.  
In both cases, formation of lunar mass objects leads to 
the production of copious amounts of dust and a significant
excess of radiation at 24 $\mu$m ($F_{24}/F_{24,0} > 1$).
As objects grow into Mars-sized or larger planets, less 
frequent collisions produce smaller amounts of dust. 
Individual large collisions are then responsible for
large increases in the dust excess.}
\end{figure}

\begin{figure}
\plotone{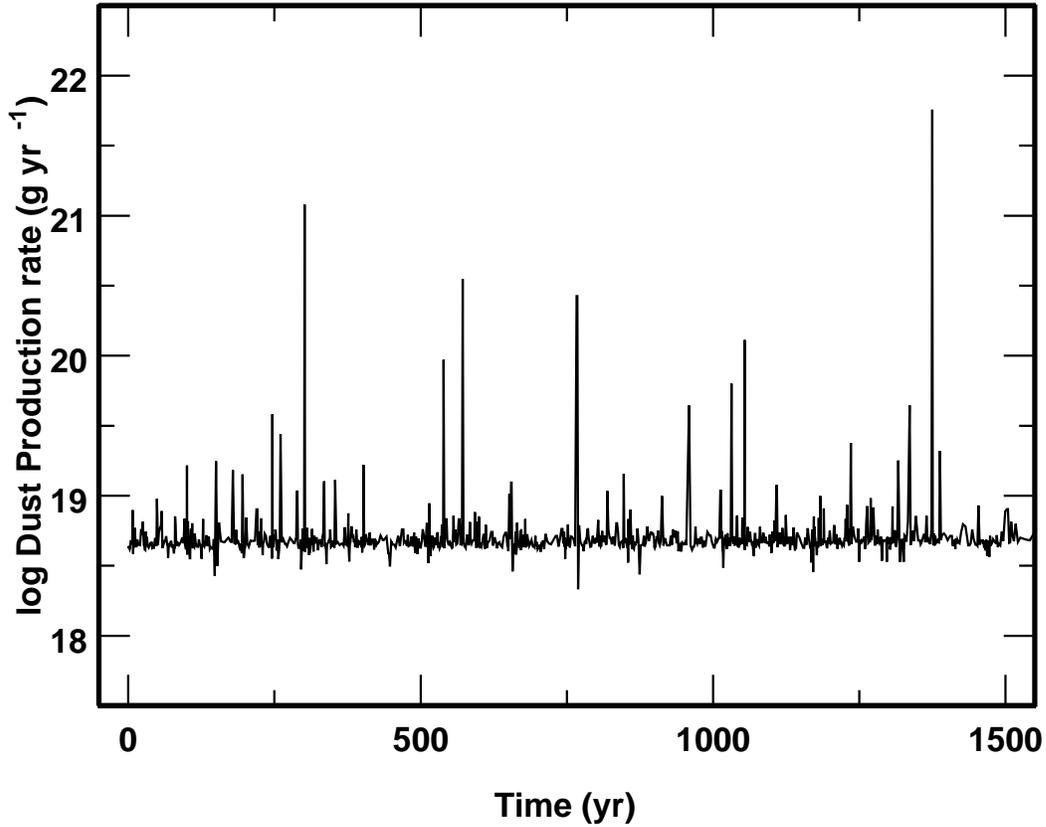} 
\caption
{Variation of the dust production rate with time for the
0.4--2 AU calculation in Figure 4 at $ 5 \times 10^6$ yr.
Numerous
collisions between 1 m to 1 km objects produce the roughly 
constant dust production rate of $\sim$ $5 \times 10^{18}$ 
g yr$^{-1}$. Binary collisions produce occasional spikes 
in the dust production rate.} 
\end{figure}

\begin{figure}
\plotone{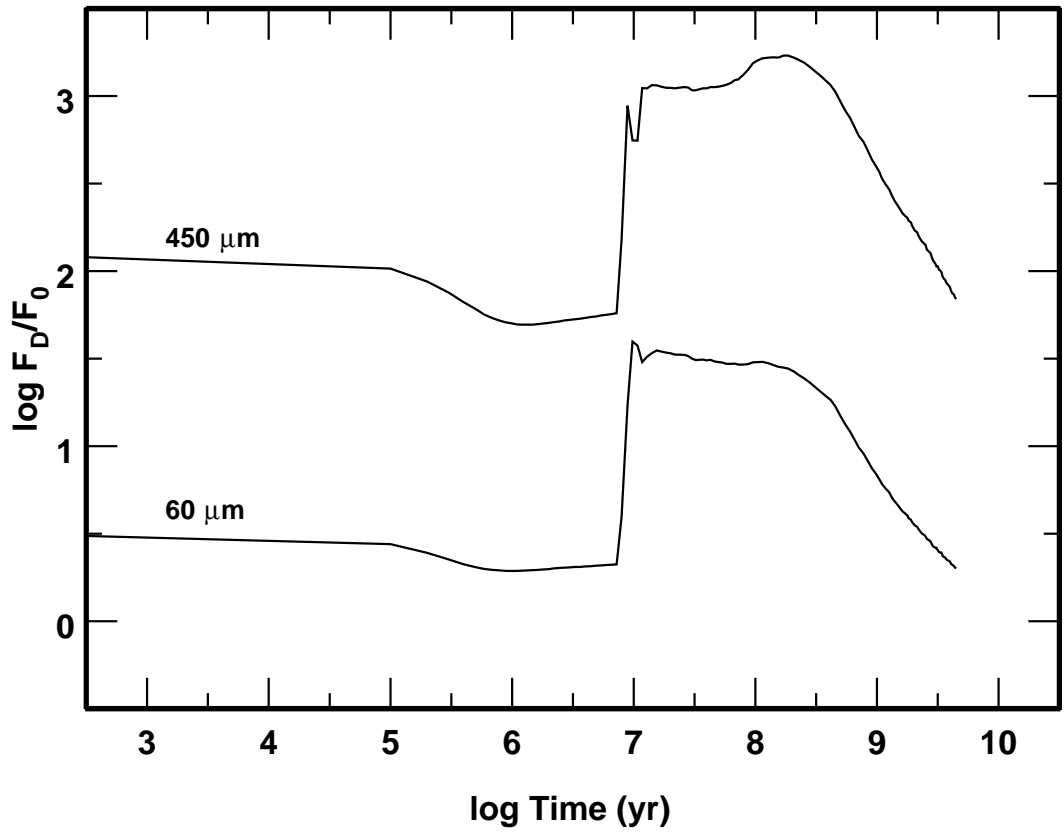} 
\caption
{Evolution of the IR excess at 60 $\mu$m and at 450 $\mu$m 
for a planetesimal disk at 30--80 AU around a solar-type star.}
\end{figure}

\begin{figure}
\plotone{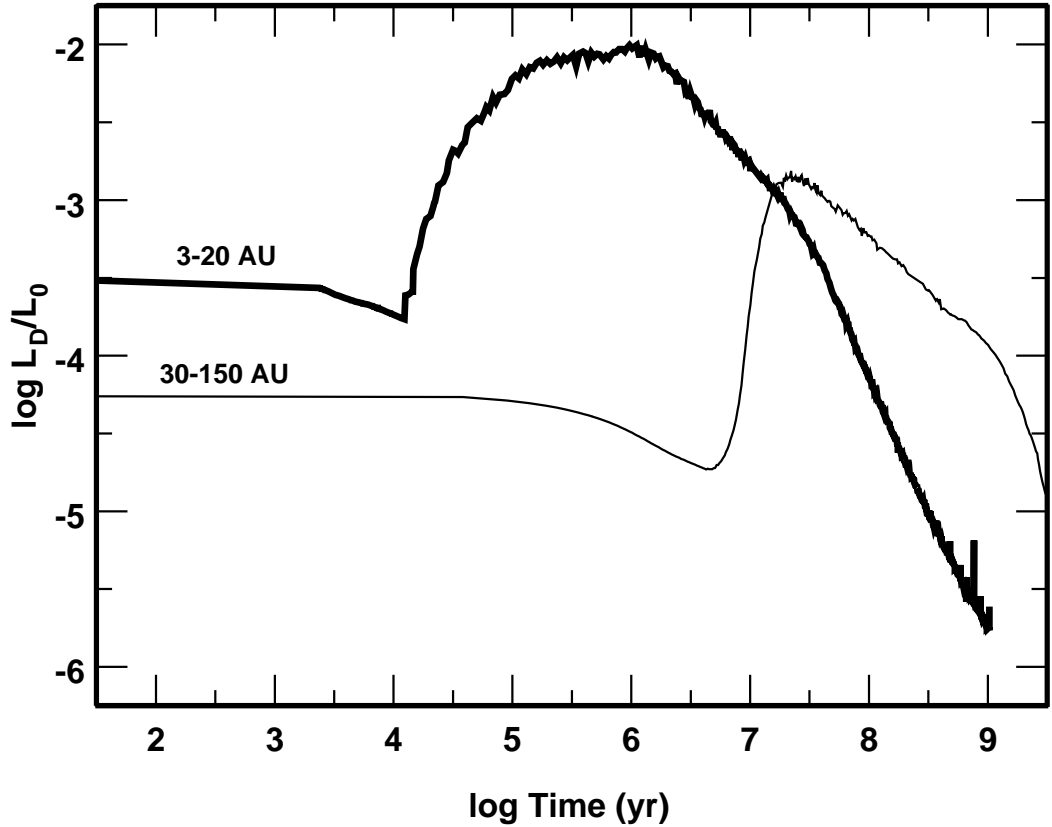} 
\caption
{Evolution of the dust luminosity for models at 3--20 AU and 
at 30--150 AU. The central A-type star has 
$M_{\star} = 3 M_{\odot}$,
$L_{\star} = 50 L_{\odot}$, and
$T_{\star}$ = 9500 K.
Binary collisions produce the small variations
in dust luminosity at the peak luminosity of each model. During
the decline, individual collisions are not visible as distinct 
luminosity spikes.}
\end{figure}

\begin{figure}
\plotone{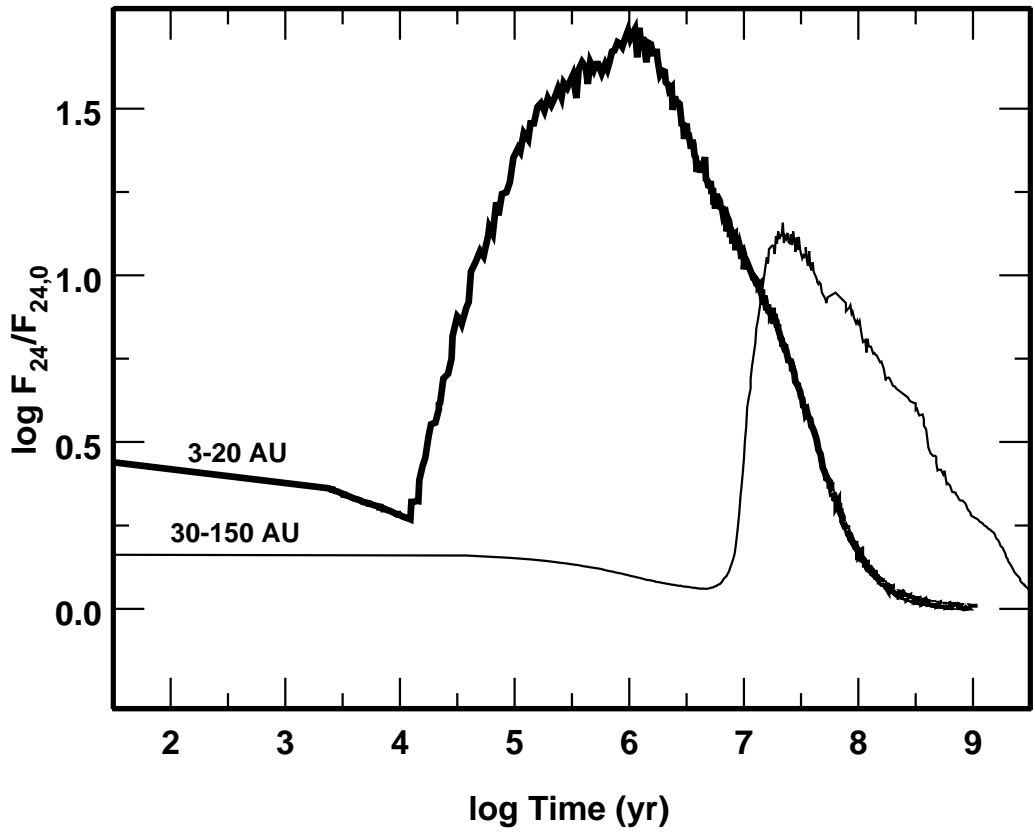} 
\caption
{As in Figure 7 for the 24 $\mu$m excess.}
\end{figure}

\begin{figure}
\plotone{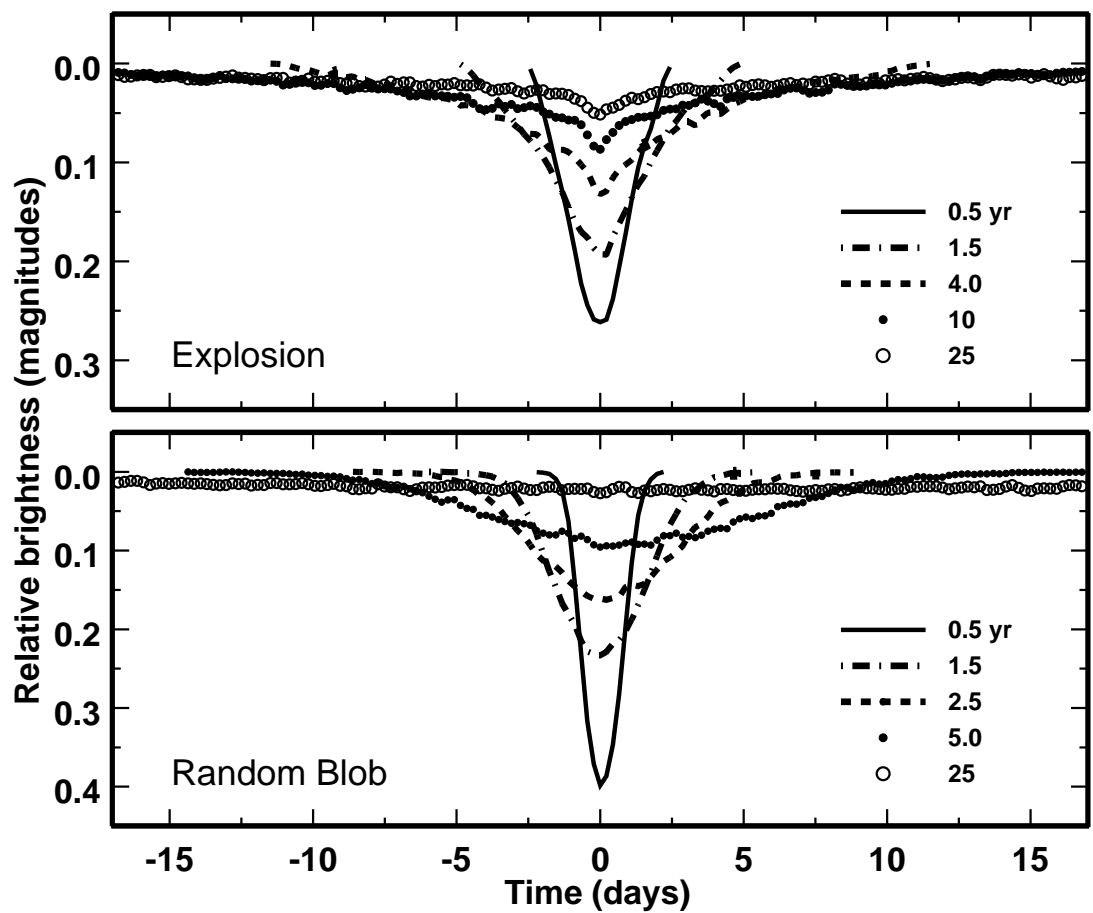} 
\caption
{Variation in brightness of a solar-type star eclipsed
by a debris cloud formed from the collision of two 1000 km
objects at 1 AU.  In these calculations, a large remnant contains
$\sim$ 15\% of the initial mass; the remaining mass is in 
smaller objects with a power-law size distribution ($q = 3.5$).
The legend in each panel indicates the time after the collision 
for each curve. Orbital shear and internal expansion of the cloud 
produce the gradual evolution in eclipse depth and duration.}
\end{figure}

\begin{figure}
\plotone{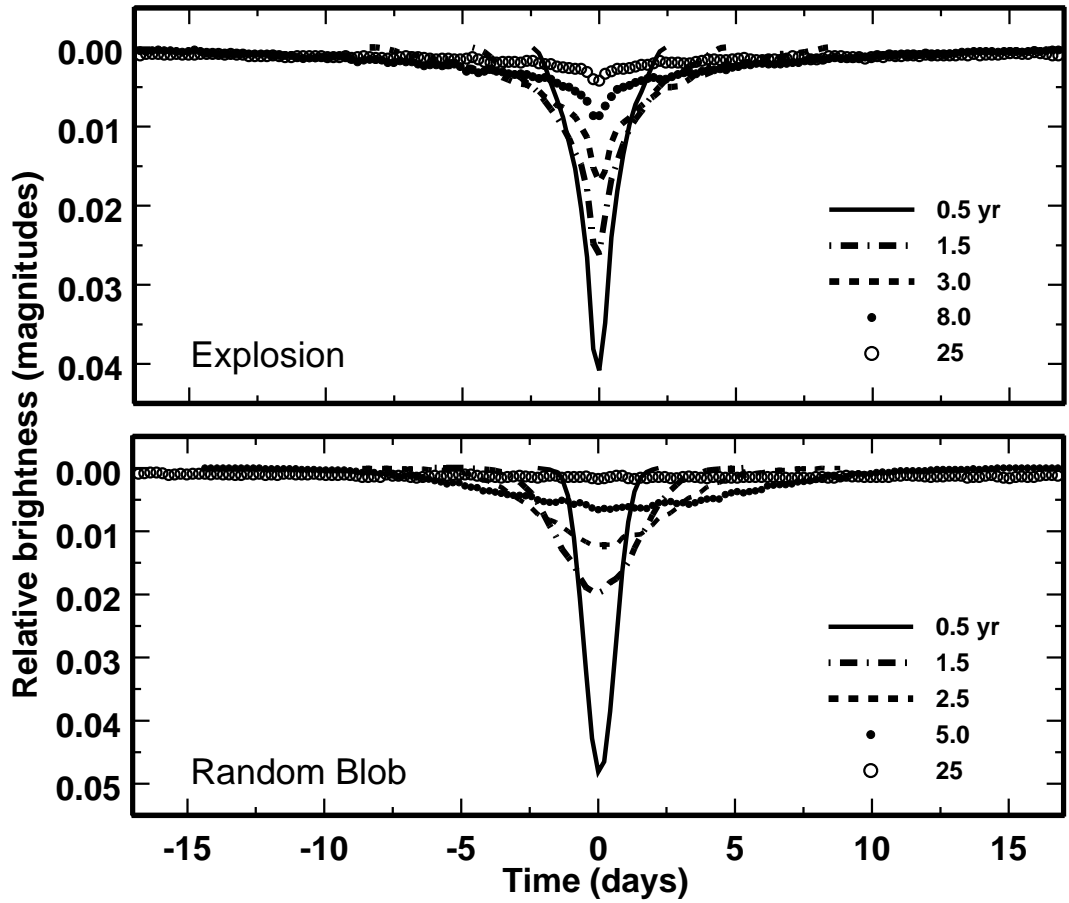} 
\caption
{As in Figure 9 for the collision of two 300 km objects.}
\end{figure}

\clearpage
\begin{figure}
\plotone{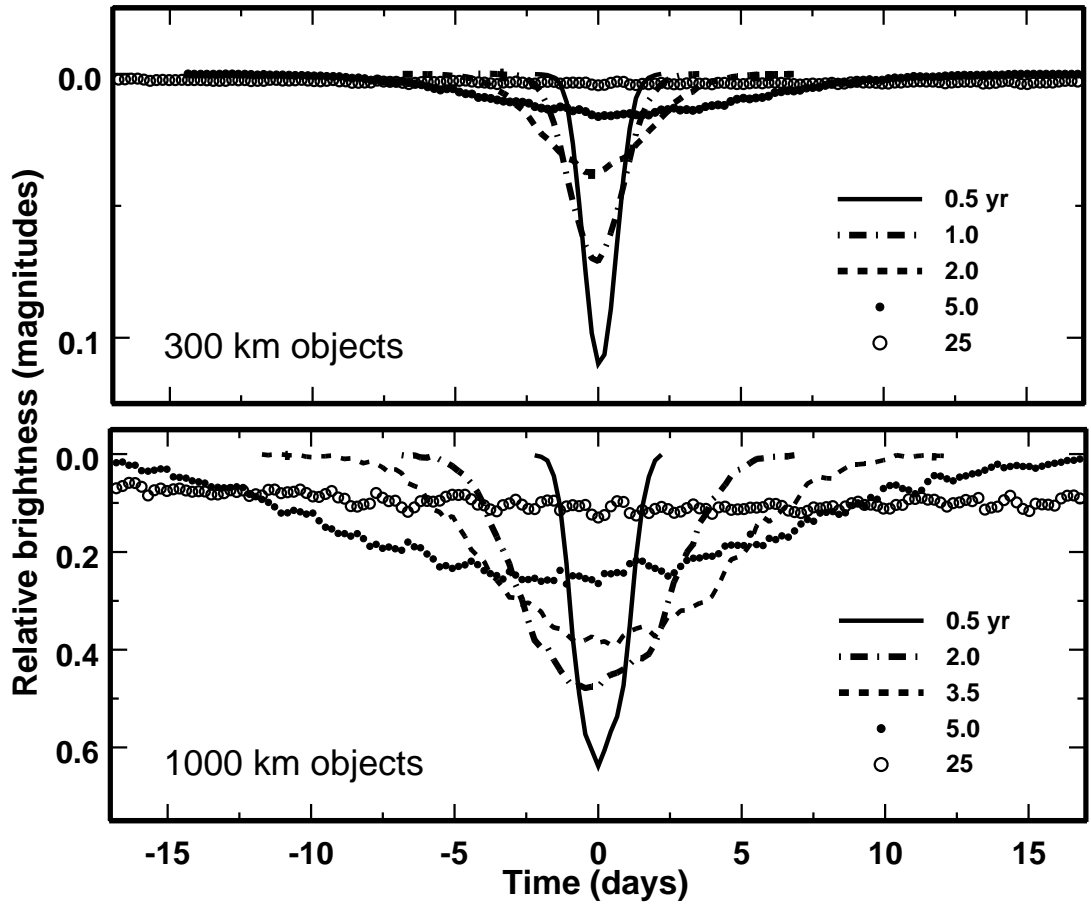} 
\caption
{As in Figures 9--10 for collisions where 20\% of the 
mass is in a single large remnant and 80\% of the mass
is in objects with radii of 10 km or smaller.}
\end{figure}

\clearpage

\begin{deluxetable}{lccccc}
\tablecolumns{6}
\tablewidth{0pc}
\tabletypesize{\normalsize}
\tablenum{1}
\tablecaption{Model Parameters}
\tablehead{
\colhead{Parameter} & \colhead{Symbol} & \multicolumn{2}{c}{G Star Models} & \multicolumn{2}{c}{A Star Models} \\
	&	& 
\colhead{A} & \colhead{B} &
\colhead{A} & \colhead{B}}
\startdata
Stellar mass ($M_{\odot}$) & $M_{\star}$ & 1 & 1 & 3 & 3 \\
Stellar luminosity ($L_{\odot}$) & $L_{\star}$ & 1 & 1 & 75 & 75 \\
Inner radius (AU) & $a_{in}$ & 0.4 & 30 & 3 & 30 \\
Outer radius (AU) & $a_{out}$ & 2.0 & 150 & 20 & 150 \\
Particle mass density (g cm$^{-3}$) & $\rho$ & 3.0 & 1.5 & 3.0 & 1.5 \\
Surface density at 1 AU (g cm$^{-2}$) & $\Sigma_0$ & 7.5--30 & 15--60 & 5--50 & 5--50 \\
Gas removal timescale (yr) & $t_g$ & $10^6$--$10^7$ & $10^7$ & $10^6$--$10^7$ & $10^7$ \\
\enddata
\end{deluxetable}

\end{document}